%% file: main.tex
\documentclass[conference,compsoc]{IEEEtran}

\usepackage{array}
\usepackage{cite}
\usepackage{epsfig,endnotes}
\usepackage{xcolor,colortbl}
\usepackage{amsmath}
\usepackage[caption=false,font=footnotesize,labelfont=sf,textfont=sf]{subfig}
\usepackage{url}

\usepackage{booktabs}
\usepackage{hhline}
\usepackage{tikz}
\usepackage{amssymb}

\usepackage[bookmarksnumbered,pdftex,pdfpagelabels=true,pagebackref=false]{hyperref} 
\hypersetup{
    plainpages=false,       
    unicode=false,          
    pdftoolbar=true,        
    pdfmenubar=false,        
    pdffitwindow=false,     
    pdfstartview={FitH},    
    pdftitle={No Privacy in the Electronics Repair Industry},    
    pdfauthor={Jason Ceci, Jonah Stegman, and Hassan Khan},    
    pdfnewwindow=true,      
    colorlinks=true,        
    linkcolor=blue,         
    citecolor=blue,        
    urlcolor=blue           
}

\newcommand{\subhead}[1]{\vspace {0.25pt}\noindent{\textbf{#1.}}}

\makeatletter
\newcommand*{\radiobutton}{%
  \@ifstar{\@radiobutton0}{\@radiobutton1}%
}
\newcommand*{\@radiobutton}[1]{%
  \begin{tikzpicture}
    \pgfmathsetlengthmacro\radius{height("X")/2}
    \draw[radius=\radius] circle;
    \ifcase#1 \fill[radius=.6*\radius] circle;\fi
  \end{tikzpicture}%
}
\makeatother
\newenvironment{SUBENVccomment}[2]{\color{#1}[#2:]~}{\color{black}}

\definecolor{author1}{rgb}     {0.9,0.5,0.0}
\definecolor{author2}{rgb}     {0.6,0.0,0.8}
\definecolor{author3}{rgb}     {1.0,0.4,0.0}

\newcommand{\new}[1]{\textcolor{black}{#1}}

\begin{document}

\title{No Privacy in the Electronics Repair Industry*}

\author{\IEEEauthorblockN{Jason Ceci, Jonah Stegman, Hassan Khan}
\IEEEauthorblockA{University of Guelph\\
\{jceci, jstegman, hassan.khan\}@uoguelph.ca
}}
\maketitle

\thispagestyle{plain}
\pagestyle{plain}

\begin{abstract}
Electronics repair and service providers offer a range of services to computing device owners across North America---from software installation to hardware repair. 
Device owners obtain these services and leave their device along with their access credentials at the mercy of technicians, which leads to privacy concerns for owners' personal data. 
We conduct a comprehensive four-part study to measure the state of privacy in the electronics repair industry. 
First, through a field study with 18 service providers, we uncover that most service providers do not have any privacy policy or controls to safeguard device owners' personal data from snooping by technicians. 
Second, we drop \emph{rigged} devices for repair at 16 service providers and collect data on widespread privacy violations by technicians, including snooping on personal data, copying data off the device, and removing tracks of snooping activities.
Third, we conduct an online survey (n=112) to collect data on customers' experiences when getting devices repaired. 
Fourth, we invite a subset of survey respondents (n=30) for semi-structured interviews to establish a deeper understanding of their experiences and identify potential solutions to curtail privacy violations by technicians. 
We apply our findings to discuss possible controls and actions different stakeholders and regulatory agencies should take to improve the state of privacy in the repair industry.


\end{abstract}

\maketitle

\input{01.introduction}

\input{02.Related_work}
\input{03.Methodology}
\input{04.Part-1}

\input{05.Part-2}

\input{06.Survey}
\input{07.Interview}
\input{08.Discussion}

\input{09.Limitation}

\input{10.Conclusion}

\section*{Acknowledgements}
This material is based upon work supported by NSERC under Grant No. RGPIN-2019-05120. 
\bibliographystyle{IEEEtranS}
\bibliography{references}
\input{11.Apendix}
\end{document}

%% file: 01.Introduction.tex
\section{Introduction}
\label{sec:introduction}


The annual global electronics repair and service industry is valued at over \$19B~\cite{ibis_report}.
Repair services are offered by a range of service providers---from mom-and-pop shops to big-box stores---to millions of users.
Geek Squad (the technology repairs and service subsidiary of Best Buy) alone serves 4.5 million customers a year~\cite{geeksquad}.
The types of services offered by these providers include software installation and troubleshooting, routine maintenance (e.g., software updates), data recovery, and hardware repair or upgrades.
Customers use these services for a variety of computing devices, including personal computers and smartphones. 
\let\thefootnote\relax\footnotetext{*Accepted at IEEE S\&P 2023}


Service or repair of a computing device often requires (administrative) access to the device's operating system for installation, diagnostics, or validation. 
During the repair, technicians have complete access to the device and may access device owners' personal data. 
This privacy violation may occur accidentally (e.g., when verifying successful recovery of a document), intentionally (e.g., snooping due to curiosity), or at the behest of law enforcement~\cite{eff_geeksquad}.
Several cases of such privacy violations have been made public through litigation and Freedom of Information Act (FOIA) requests. 
Electronic Frontier Foundation has uncovered that Geek Squad employees have been invading device owners' privacy at the behest of the FBI, potentially circumventing device owners' Fourth Amendment rights~\cite{eff_geeksquad}.
In addition, individuals have reported the theft of their nude pictures by Apple store~\cite{apple_snooping} and Geek Squad~\cite{geeksquad_snooping} technicians. In one instance, employees at a smartphone repair service stole and distributed nude images for over seven years~\cite{mobile_snooping}.

While it has been over a decade since the first incident was publicly reported~\cite{geeksquad_snooping}, policies and controls to safeguard customers' data are still \new{inadequate}. 
Geek Squad provides a privacy policy for device repairs in the US on their website, which states, ``[O]ur Geek Squad Agents are trained to never access data on a customer's device provided to Geek Squad for service except in limited circumstances, and only to the extent necessary to perform the service, such when you ask us to recover your data''~\cite{bestbuy_privacy}.
However, no information is available on the controls employed to implement this policy. 
Similarly, after an incident in 2016 at an Apple Store, Apple said in a statement: ``When we learned of this egregious violation of our policies at one of our vendors in 2016, we took immediate action and have since continued to strengthen our vendor protocols'', only for another incident to be reported in 2019~\cite{bbc_apple_snooping}.
If technicians get caught snooping, they are terminated from their jobs~\cite{apple_snooping} and possibly face legal proceedings~\cite{mobile_snooping}. However, these reactionary measures fail to treat the underlying problem.  

The electronics repair industry provides economic and environmental benefits. However, there is a dire need to measure the current privacy practices in the industry, understand customers' perspectives, and build effective controls that protect customers' privacy.
To this end, we conduct a carefully designed four-part study, which involves both stakeholders---\new{electronic repair service providers (\emph{``service providers''})} and customers. 
The first two parts of the study focus on the service providers for which we target big-box stores, regional chains, and local mom-and-pop stores across three cities in Canada.
In the first part, researchers dropped devices for a battery replacement service (a service that does not require access to the device OS) to measure the availability and communication of privacy policy and requests for device access credentials. 
During the second part, we \emph{rigged} devices to log all the actions performed by the user. These devices were then dropped off  with their audio driver disabled to get the audio issue fixed. After repairs were completed, the logs were analyzed to note the type of privacy violations perpetrated by the technicians.
The third part was an online survey (n=112), which collected users' perceptions and experiences around device repairs.
In the final part, a subset of interview participants (n=30) was invited to develop deeper insights into their experiences and to explore possible ways to improve customers' privacy when their devices are being repaired. 

To the best of our knowledge, our work provides the first-ever holistic view of customers' privacy in the electronics repair industry. Our key findings include:\looseness=-1

\begin{itemize}

    \item Privacy policies and the practice of communicating protocols and controls to protect customers' data do not exist across service providers of all sizes. 
    \item Service providers \new{largely (10/11)} require ``all-access'' to the device, even when it is not needed.
    \item Technicians often snoop on customers' data (6/16) and sometimes copy those to external devices (2/16). 
    \item Technicians who violate privacy \new{often} do so carefully not to \new{generate evidence (1/6) or remove such evidence (3/6)}. 
    \item A significant proportion of broken devices (26/79 (33\%)) are not repaired due to privacy concerns. For the devices that get repaired, device owners are concerned about threats to their privacy but do not use the proper controls to protect their data.

\end{itemize}
We outline policy improvements, possible controls, and the actions and roles different stakeholders and regulatory agencies need to play to improve the state of privacy in the repair industry.


%% file: 02.Related_work.tex
\section{Related Work}
\label{sec:related_work}
Researchers have explored privacy concerns of device owners when sharing their devices temporarily with acquaintances~\cite{hang2012too,mazurek2010access,matthews2016she,karlson2009can} and when there is unauthorized  physical access to personal devices by acquaintances~\cite{marques2019vulnerability}. Existing security controls are available that enable secure temporary device sharing (e.g., guest accounts~\cite{androidguestaccounts}).
However, these security controls are not designed to protect from  technicians, who often need administrative access to the device. 
Researchers and regulatory authorities have uncovered poor privacy practices at electronic retailers when refurbishing used devices~\cite{opc_report, ceci2021concerned}. 
While these retailers may be providing repair services, the customer's intent is not to get the device repaired.
Some researchers have explored the knowledge and skills of technicians~\cite{jackson2012repair, jackson2014breakdown} and how the repair infrastructures are set up in \new{low income} countries~\cite{jang2018crowdsourcing, jang2019trust}.
However, these works do not explore the customers' privacy aspects. 

While news articles abound on privacy violations (see \S~\ref{sec:introduction}), the state of customers' privacy in the repair industry has not received a holistic academic treatment.
The only work on this subject is from Ahmed et al.~\cite{ahmed2016privacy}, where the authors' goal was to ``improve the practices surrounding the repair of digital artifacts in developing countries.'' The authors conducted this study in Bangladesh. Through an ethnographic study,  a researcher interned at a smartphone repair shop for three months to note the technician and customer interactions.
This was coupled with an online survey and semi-structured interview to understand customers' privacy concerns. They found that technicians often snoop and sell the personal data of customers. Furthermore, while some customers reported their suspicions of possible privacy violations, a lack of understanding around privacy and repair was common among participants. Finally, they collected feedback on an Android application, which flagged privacy violations by logging which apps were used during repair. 

Ahmed et al.'s investigation focuses on a \new{low or middle income} country and privacy in the context of smartphones repairs only.  There are considerable differences between \new{high and low income} countries when it comes to the use of technology~\cite{sawaya2017self, vashistha2018examining, redmiles2019should} and privacy expectations and regulations~\cite{mare2017security, ahmed2016privacy}.
With the media coverage and subsequent promises from some service providers in the US~\cite{bbc_apple_snooping,bestbuy_privacy}, in addition to measuring customers' perceptions, it is important to understand the existence and communication of policies and controls to protect customers' privacy. 


%% file: 03.Methodology.tex
\section{Study Design}
\label{sec:design}

\new{We aim to investigate the following questions:}
\begin{enumerate}\itemsep0em
    \item[RQ1] Does the electronic repair industry have privacy policies or procedures to safeguard customers' data? If so, how are those policies or procedures communicated to customers?
    \item[RQ2] Do repair service providers only request access to resources that are needed?
    \item[RQ3] Do service technicians access customers' data? If so, how widespread is this issue, and what type of violations are common?
    \item[RQ4] What are customers' understanding of risks around device repairs, and how does it influence their device repair decisions and repair preparations?
    \item[RQ5] What solutions are viable to improve the state of privacy in the electronics repair industry? 
\end{enumerate}

There are several challenges to our investigation.
Service providers are of various sizes and different service providers may have different maturity levels and corporate oversight. 
A holistic investigation should investigate these research questions across different service providers.
Moreover, customers may need different types of repairs or services. Some services may only need access to the device hardware, while some services may require access to the device OS. The type of access required may change for different device types for the same service. To collect meaningful data, it is important to carefully choose the appropriate device and service. 

To find answers to these questions, we conducted a four-part study.
The first part of the study collected data on RQ1 and RQ2 by initiating repairs at various service providers. 
Note that the first part required the investigator to ask the technicians about the existence of privacy policies. This may have influenced the technicians' interaction with the device. 
Therefore, to investigate RQ3, we collected data using a separate study. For the second part, we had \emph{rigged} devices repaired that logged all the interactions of technicians during the repair process.
Next, we conducted an online survey (n=112) for RQ4, which measured the respondents' understanding of risk perceptions, how it influenced their repair decision, safeguards they employed when getting the device repaired, and any privacy violations they noticed for the repair services that they requested. The findings were enriched, and RQ5 was investigated through a follow-up interview with a subset of survey participants (n=30).
We document the recruitment process, study procedure, and results separately for the four parts in their respective sections.   

\subhead{Controlling Confounding Factors}
We used various measures to control confounding factors.
For different parts of the study, an investigator never revisited the same service provider.
The investigators did not provide their real names at the device drop-off in case the technicians searched their names on the Internet to uncover that the investigators were security researchers.
We used different devices for different branches of service providers and different devices for different parts of the study to ensure that a record of the device did not exist in the service providers' central database.
Finally, the advertisement and recruitment for the survey and interview strictly began after the conclusion of the first two parts. This order ensured that we did not influence the behaviour of the technicians due to the knowledge of this study.\looseness=-1 

\subhead{Choosing Repair Service Providers}
The first two parts of the study require interactions with the service provider. 
We divided service providers into three categories. 
A \emph{``National''} service provider is a big-box store that is a subsidiary of a publicly traded organization across North America.
A \emph{``Regional''} service provider is a store of a larger chain that has stores across more than two cities.
A \emph{``Local''} service provider is a  service provider local to the city,  is operating at a commercial location (i.e., not operating out of their home) and has a business license. When choosing local service providers, we only selected those service providers who had a website and a Google Maps rating of 3.5 stars or higher.

\subhead{Ethical Considerations}
\new{
We received approval from our institutional IRB for all four parts of the study. Part-3 and Part-4 required informed consent, PII anonymization, and allowing participants to withdraw their data up to two weeks after the study. 
Part-1 and Part-2 required covert observations to study the policies and procedures of the service providers, which are \emph{commercial entities}. We did not collect any personally identifiable information about the technicians or representatives. To avoid influencing the behaviour of the technicians, we did not obtain informed consent, Part-1 required limited disclosure, and Part-2 required deception.  
 We strictly adhered to the guidelines in Canada's Tri-Council Policy Statement (TCPS-2)~\cite{tcps2} for covert observation studies where informed consent could not be obtained.  To this end, several necessary measures were used to address privacy and confidentiality issues. 
 To protect the privacy of the repair technicians, we securely deleted all data where the technician may have used the device to view personal data (e.g., accessing a personal website or personal file on an external drive). 
 Second, all location tracking and additional information collection in the OS were disabled. After each experiment, only logs relating to privacy violations and covering tracks (e.g. removing logs or clearing ``recent files'') were retrieved before reimaging the device. 
 Third, we anonymized the collected data and only report the type of service provider (national, regional, or local) without disclosing which of the three cities the service provider belonged to. 
}

\new{
We chose to not debrief the service providers for several reasons (as permitted by TCPS-2). 
Previous incidents of technicians' misconduct have resulted in the technician losing their job. 
Informing the service providers that they were a part of this study, regardless of anonymization assurances, may cause them to feel threatened and potentially retaliate. 
Our experiment was designed to pose minimal risk to the technician and their privacy. The technicians were paid for their services and there is no expectation of privacy for the repair actions taken on the experimenter's device (we do not disclose any repair procedures). Given the minimal risks to technicians, a debrief provides little advantage. 
On the contrary, disclosing the study to technicians who may have committed privacy violations may lead to psychological and social harms including feeling ashamed, a fear of social stigma, and guilt. There are precedents~\cite{finn2007designing, smith1975faith, chiou2012new} on not debriefing participants in technology and social science research for these reasons. For instance, Finn and Jakobsson did not debrief participants of their phishing experiment as ``[The subjects] may feel upset, anxious, or angry that they were fooled''~\cite{finn2007designing}. Interested readers are referred to the work by Roulet et al.~\cite{roulet2017reconsidering} for a discussion on consent and debriefing in covert participation experiments and several precedents in the social sciences domain.  
While unconventional, our experiment design enables us to conduct legitimate privacy research in the public interest, which was otherwise not possible and which has the potential to improve the state of privacy, particularly for novice users who often engage with these service providers.}

%% file: 04.Part-1.tex
\section{Part-1: Privacy Policies and Protocols}
\label{sec:part1}
 The first part of the study was conducted from August 2021 to September 2021. The main objectives were to measure: (1) the existence and communication of privacy policies to safeguard customers' data; 
and (2) what information is collected for a repair procedure that does not require access to the device OS.

\subsection{Procedure}
To study the privacy policies and protocols, we visited eleven service providers---three national, three regional, and five local service providers. We also report our findings for two national smartphone repair service providers and for five device manufacturers. 
We only visited one store for the national and regional service providers as we expected the existence and communication of policies to be the same across different stores of the same service provider. 
For the repair service, we chose battery replacement for Asus UX330U laptops as its battery replacement only required removal of the back of the device and the device BIOS provided all the details about the battery health. Therefore, it can be argued that the technicians did not need access to the device OS for diagnosis, replacement, or verification of the repair. 
Understandably, several repair or service procedures do require access to the device OS, which leaves no option for the customer but to provide their credentials. For such repair procedures, our findings on the safeguards adopted by the service providers are of relevance. 

A script was prepared for the drop-off. 
The researcher was required to look for any notice that outlined the privacy policy and take a picture of it right before leaving. 
The researcher was then required to explain the issue (``the battery is dying and needs replacement'') and make a mental note of the technician's requests and questions. If the technician asked for credentials, the researcher asked them why it was needed and how customers' data would be protected.  After providing the credential, the researcher noted how it was stored. Finally, the researcher asked about any privacy policies and protocols  to safeguard customers' data (``how do you make sure no one will access my personal data?''). Within ten minutes after the drop-off, the researcher completed a survey documenting their experience (see Appendix~\ref{app:reconnaissance_survey}).

\subsection{Results}
\label{subsec:part1_results}
For qualitative analysis of technicians' responses (e.g., response to ``how customers' data was protected''), we performed thematic analysis to identify themes (see details on thematic analysis in \S~\ref{subsec:interview_results}). The quotes from the technician were recalled by the researcher and are not verbatim. 

\subhead{Privacy Policies}
None of the service providers posted any notice informing customers about their privacy policies. Similarly, until the devices were handed over, no researcher was informed about a privacy policy, their rights as a customer, or how to protect their data. 
After the devices were handed over, only three national and three regional service providers  
provided a terms and conditions document for the customer's signature. 
However, these terms and conditions only communicated that the service provider was not liable for the loss of the customer's data and that the customer was responsible for backing up their data. No document listed any safeguards on how customers' data was protected, or the steps customers should take to safeguard their data from snooping. 
All three national service providers provided a link to their online privacy policy, which was a generic policy on data collection and retention (i.e., not specific to electronics repair services). 


\subhead{Access to Credentials}
All but one repair service provider asked for the credentials to the device OS---a regional service provider was the only exception. 
When technicians asked for the credentials, the researcher asked why it was needed. 
All three national service providers \emph{needed} the credentials for repairs regardless of the service as it was a part of the ``paperwork'' (\textit{``We need the password for all repairs.''}  (National-1)).
Technicians from one local and two regional service providers said that they needed the credentials to run diagnostics. 
Technicians from four local service providers said that they needed it for the verification of the performed service. 

We asked the technicians if the device battery could be replaced without providing the credentials. 
The three national service providers could not take the device without the credentials. 
Two regional and two local service providers agreed to do it but informed the researcher that they would not be able to verify their work and would not be responsible for the quality of their work. 
Finally, one local service provider asked the researcher to remove the password if they did not want to share it, and one local service provider told the researcher that if the password was not provided, they would reset the device if required.

\subhead{Storage of Access Credentials}
Users widely reuse their credentials~\cite{das2014tangled}. It is possible that the password shared with the technicians  is reused elsewhere~\cite{khan2020widely}. 
For the repair service, the customer has to provide the password along with their name, contact number, and email address.
After providing this information to the technicians, the researcher noted how this information was stored. 
All three national service providers stored this electronically in a database. No information was shared with the customer on who will have access to this information or how long it will be stored.
The three regional service providers stored this information in an electronic database too. Furthermore, two of these service providers printed a label with the customer's name, contact number, and password. This label was pasted on the device and its charger, which was accessible to all staff members. 
Finally, for the local service providers, 3/5 put a sticky note with the password on the device itself and 2/5 stored this information in an electronic database.





\subhead{Protection of Personal Data}
The researcher asked the technician if anyone would access the personal data on the device during the repair. All ten service providers who obtained credentials responded that the personal data would not be accessed, and only the technician working on the problem would access the device. 
We asked the technicians how they ensure no one would access data on the device.
Their responses conveyed no indication of a protocol or control to protect access to customers' data. 
Their responses were codified \new{and the inter-rater agreement between the two researchers was moderate (Fleiss's $\kappa$ = 0.48)}. 
6/10 service providers (including two national service providers) ``reassured'' the researcher. Their responses were similar to: \textit{``The technicians only do the repair and nothing else.''} (National-1). 
The response from 4/10 repair service providers was unsatisfactory or a deflection. Two representative responses are provided below:
\begin{Q}
\textit{``It will be sent to a technician elsewhere.''}  (Regional-1)
\end{Q}
\begin{Q}
\textit{``It is an honour system. We wouldn't be in business if we were doing that.''}  (Local-4)
\end{Q}


\subhead{Smartphone Repairs and Device Manufacturers}
Most service providers that we investigated for this part of the study repaired both smartphones and computers. We also approached two regional smartphone repair service providers (who only repaired smartphones) to replace the battery of a Samsung S8 smartphone.
Similar to our findings for laptops, the researchers found no notice or communication on the privacy policy and how their data will be protected. One service provider requested the credentials, and the other did not. When the service provider was asked why the credentials were needed, they informed the researcher that it was needed to run the ``Samsung Diagnostic utility.'' When asked how this data would be protected, the service provider reassured the researcher that no one would look at customers' personal data. 
An interesting aspect to note is that the terms of service of this regional service provider contradicted their reassurance:
\begin{Q}
\textit{``[Redacted name] will not treat data on your device as confidential and disclaims any agreement with you or other obligation to do so.''}  (Regional-4)
\end{Q}


We also contacted six laptop manufacturers and requested a repair for their brand of devices. 
One manufacturer had outsourced repairs to two national service providers, so we exclude it from the following results.
The mail-in instructions that the manufacturers provided were studied to see their policies on safeguarding customers' data and collection of credentials.  
All device manufacturers stated that they had the right to delete all data on the device.
 4/5 manufacturers suggested that the user should back up their device prior to mailing. 
 All generic mail-in instructions required credentials and contained warnings to the effect of:
 \vspace{-1em}
 \begin{Q}
\textit{``Failure to provide your username and password may delay or prevent us from completing your repair. You may also remove the system password prior to shipping the unit instead of providing it above.''} (Manufacturer-2) 
\end{Q}
Such statements from the manufacturers leave little room for those customers whose device repair could be completed without access to the device OS. 
Finally, none of the mail-in instructions listed any mechanisms that customers could use to safeguard their sensitive data before sending the device.

%% file: 05.Part-2.tex
\section{Part-2: Privacy Violations During Repair}
\label{study2}
The second part of the study was conducted from October 2021 to December 2021. The purpose of the second part was to collect data on privacy violations during repairs. There are several factors that influence this data collection that are not controllable. First, a technician with the intent to snoop may not get the opportunity to do so due to the presence of other individuals around them or deadlines. Second, only select technicians may be involved in snooping, and they may not get the device dropped off by the researcher for the repair service. Consequently, to concretely establish that privacy is ``regularly'' violated at a service provider, we need to collect a large number of samples. Since this requirement has obvious logistical issues, we only present our results for a limited number of device repairs.

\subsection{Methodology}
\label{sec:2-methodology}
We bought new laptops with Microsoft Windows 10 and disabled the audio driver on these laptops. We chose the audio issue as: (1) it was a simple and cheap repair; (2) it did not require access to users' files (unlike software installation or Virus removal); and (3) the problem was easy to create (i.e., disabling the driver). 
A total of six devices were used in this experiment. Three devices were configured with a male persona, and three were configured with a female persona.
We set up email and gaming accounts, and populated browser history across several weeks. We added documents, pictures from the experimenters, and a cryptocurrency wallet with credentials. 

\new{We also added female-coded revealing pictures. Female-coded pictures were used as most of the recently reported privacy violations were committed against females~\cite{apple_snooping, mobile_snooping,geeksquad_snooping}. Furthermore, researchers have shown that female and non-binary individuals are more likely to face issues from non-consensual image sharing, like a technician accessing devices~\cite{lenhart2016nonconsensual,geeng2020usable}. 
We reached out to verified Reddit posters on the ``r/GoneMild/'' subreddit, where posters post revealing pictures without any nudity. 
 Other researchers \cite{vandernagel2013faceless, shelton2015online, matzen2017streetstyle} have previously used and published pictures from the ``r/GoneWild/'' subreddit and discussed the best practices to ensure posters' anonymity~\cite{vandernagel2013faceless}. We followed these practices and shortlisted those posters who were verified, and whose posts did not contain their face, identifying features (tattoos, backgrounds), or any watermarks.
Two out of seven posters responded to our message and gave us permission to use their photos  ``for research with the possibility that their photos could be viewed or copied by a computer technician operating in Ontario, Canada''.
 The names and metadata of the images were scrubbed before use.
}\looseness=-1

We appropriately tweaked the content to reflect the male or female personas. 
After preparing the devices, they were imaged, and these images were deployed before each drop-off. We developed a logging utility which is a wrapper on the Windows Problem Steps Recorder~\cite{psr}. Windows Problem Steps Recorder can execute in the background (through CLI) and captures the screen on every mouse click. It also records the keys pressed by the user. Our logging utility executed in the background as a Windows process and logged all interactions (referred to as ``interaction logs'') of the technicians. In addition, we enabled Windows Audit Policy to log access to any file on the device.

We targeted two national, two regional, and eight local service providers.
We considered two branches for each national and regional service provider. 
The repair requests were gender-balanced. 

 \begin{table}[t]
 \caption{Recorded Privacy Violations. Each Symbol Represents Violations in a Unique Experiment. (\textbf{F}=Female; \textbf{M}=Male)}
 \scalebox{0.80}{
\begin{tabular}{lllllll|}
\cline{2-7}
\multicolumn{1}{l|}{}                             & \multicolumn{2}{c|}{\textbf{National}}                  & \multicolumn{2}{c|}{\textbf{Regional}}                  & \multicolumn{2}{c|}{\textbf{Local}} \\ \cline{2-7} 
\multicolumn{1}{l|}{}                             & \multicolumn{1}{l|}{F x 2} & \multicolumn{1}{l|}{M x 2} & \multicolumn{1}{l|}{F x 2} & \multicolumn{1}{l|}{M x 2} & \multicolumn{1}{l|}{F x 4}  & M x 4 \\ \hline
\multicolumn{7}{|l|}{\textbf{Privacy Violation}}                                                                                                                                                            \\ \hline \hline
\multicolumn{1}{|l|}{Accessed Documents Folder}        & \multicolumn{1}{l|}{}      & \multicolumn{1}{l|}{}      & \multicolumn{1}{l|}{$\star$}      & \multicolumn{1}{l|}{$\diamond$}      & \multicolumn{1}{l|}{$\dagger \ddagger$}       &       \\ \hline
\multicolumn{1}{|l|}{Accessed any Picture Folder}    & \multicolumn{1}{l|}{$\ast$}      & \multicolumn{1}{l|}{}      & \multicolumn{1}{l|}{$\star$}      & \multicolumn{1}{l|}{$\diamond$}      & \multicolumn{1}{l|}{$\dagger \ddagger$}       &       \\ \hline
\multicolumn{1}{|l|}{Accessed Revealing Pictures} & \multicolumn{1}{l|}{$\ast$}      & \multicolumn{1}{l|}{}      & \multicolumn{1}{l|}{$\star$}      & \multicolumn{1}{l|}{$\diamond$}      & \multicolumn{1}{l|}{$\dagger \ddagger$}       &       \\ \hline
\multicolumn{1}{|l|}{Accessed Finance Folder}     & \multicolumn{1}{l|}{}      & \multicolumn{1}{l|}{}      & \multicolumn{1}{l|}{}      & \multicolumn{1}{l|}{}      & \multicolumn{1}{l|}{$\dagger$}       &       \\ \hline
\multicolumn{1}{|l|}{Accessed Browsing History}   & \multicolumn{1}{l|}{}      & \multicolumn{1}{l|}{}      & \multicolumn{1}{l|}{}      & \multicolumn{1}{l|}{$\diamond$}      & \multicolumn{1}{l|}{}       &     $\triangleright$   \\ \hline
\multicolumn{1}{|l|}{Copied Customers' Data}      & \multicolumn{1}{l|}{}      & \multicolumn{1}{l|}{}      & \multicolumn{1}{l|}{}      & \multicolumn{1}{l|}{$\diamond$}      & \multicolumn{1}{l|}{$\dagger$}       &       \\ \hline
\multicolumn{7}{|l|}{\textbf{Covering Tracks}}                                                                                                                                                              \\ \hline \hline
\multicolumn{1}{|l|}{Cleared ``Quick Access''}        & \multicolumn{1}{l|}{}      & \multicolumn{1}{l|}{}      & \multicolumn{1}{l|}{}      & \multicolumn{1}{l|}{$\diamond$}      & \multicolumn{1}{l|}{$\dagger \ddagger $}       &       \\ \hline
\multicolumn{1}{|l|}{Cleared Logs}                & \multicolumn{1}{l|}{}      & \multicolumn{1}{l|}{}      & \multicolumn{1}{l|}{}      & \multicolumn{1}{l|}{}      & \multicolumn{1}{l|}{$\bullet$}       &    $\triangleleft$   \\ \hline
\end{tabular}
}
\label{tab:violations}
\end{table}

 \subsection{Results}
 The devices were collected after repair, and the audit and interaction logs were copied to an external storage device for analysis.
 Except for two service providers, every provider kept the device overnight. (All national service providers kept the device for at least two days.) 
 Two service providers (one regional and one local) asked the experimenters to wait briefly and diagnosed and fixed the issue in front of the experimenters.  
 We were unable to extract logs for two local service providers (one male and one female experimenter). While we could not find a plausible reason for one of them, the other told us that they installed antivirus software and disk cleanup to \emph{``remove multiple viruses on the device}.'' (The devices were new, re-imaged between drop-offs and were not infected.) 
 Therefore, snooping detection was only possible for four national, three regional, and five local service providers.

\subsubsection{Privacy Violations} 
For our analysis, we categorized privacy violations into six categories: accessing users' data folder (containing documents), any of the picture folders, revealing pictures, finance folder, browsing history, and copying users' personal data to an external storage device. We note that other types of violations are possible, but we only report observed violations. Table~\ref{tab:violations} shows privacy violations for the two types of experimenters. 
We only noted one violation from one national service provider against a female experimenter. The folders that contained pictures and revealing pictures were accessed.
For regional service providers, we noted one violation each against male and female experimenters. 
The documents, pictures, and revealing pictures were accessed for both experimenters. 
The browser history of the male experimenter was also viewed by the technician, and the revealing pictures were zipped and transferred to an external storage device.
For the local service providers, we only note one violation against the male experimenter (browser history was accessed) and two violations against the female experimenter. The technician at one local service provider accessed documents, pictures, and revealing pictures. The technician at the other local service provider committed all violations except viewing the browsing history. The technician also copied a password-containing file and the revealing pictures to an external device.\looseness=-1


\subsubsection{Covering Tracks} 
Our logs show that after privacy violations, some service providers cleared their tracks by clearing items in the ``Quick Access'' or ``Recently Accessed Files'' on Microsoft Windows.
This behaviour was observed for three service providers (one regional against the male experimenter and two local service providers against the female experimenter).
Table~\ref{tab:violations} shows that a national service provider and a regional service provider committed privacy violations but did not cover their tracks. The interaction logs show that when viewing pictures, the technicians zoomed in on thumbnails. Consequently, these violations never left a trace (i.e., appeared in the list of recently accessed files). 



%% file: 06.Survey.tex
\section{Part-3: Online Survey}
\label{sec:survey}
The primary objective of the online survey was to understand the electronics repair industry from the customer's lens. 
We collected data on any repairs that customers got done in the past five years for two categories of devices: 1) smartphones or tablets; and 2) laptops or desktop computers. 
These two categories were chosen as these categories contain large amounts of users' personal data (during the interview, we investigated repairs for other device types too).
If respondents got devices repaired for both categories, we collected responses for both categories. 
We limited responses to the past five years for a more accurate recall. 
Where applicable, we solicited feedback on the most recent repair.
The survey was piloted with four participants, and their feedback was used to revise the survey.

\subsection{Recruitment and Procedure}
We recruited respondents for a 10-minute online survey by placing advertisements on Kijiji and Facebook, and through word-of-mouth. 
We used  Qualtrics for the survey (provided in Appendix~\ref{app:online_survey}). 
The survey collected data from respondents for the following data categories: 
(1) demographics and background; (2) devices needing repairs and the reasons for not getting repairs; (3) who repaired the device and what protocol was followed for repair services; and (4) safeguards employed by the respondents and the service providers to protect their personal data. 
At the end of the survey, respondents were asked if they wished to be contacted for a follow-up interview. Respondents were compensated \$5 for their participation.

\subsection{Results}
For test statistics, we use Pearson's Chi-Squared test to compare categorical data and Kruskal-Wallis one-way analysis of variance to compare Likert scale responses between respondents (e.g., their technology proficiency levels). 

 \subsubsection{Demographics and Background}
The survey was completed by 112 respondents (26 incomplete responses were filtered). \new{We did not allow multiple submissions from the same device and excluded responses that had failed the CAPTCHA challenge.}
Respondents were asked to provide their age, gender, and level of proficiency in technology. 
44\% (49/112) of respondents were female, and 54\% (60/112) were male. \new{In terms of their age, 29\% (33/112) reported between 18--25 years, 27\% (30/112) between 26--30 years, 19\% (21/112) between 31--35 years, 12\% (13/112) between 36--40 years, 4\% (5/112) between 41--45 years, 3\% (3/112) between 46--50 years, and 6\% (7/112) over 50 years old.}  
While 56\%  (63/112) of respondents are 30 years of age or younger, we have a good representation of other age groups. Fewer respondents self-reported a basic proficiency in technology (16\% (18/112)) compared to those who self-reported as intermediate (64\% (72/112)) or advanced (20\% (22/112)). Given that the study was advertised and conducted online, this smaller proportion is somewhat expected.

\subsubsection{Broken Devices}
\begin{figure}[t]
    \centering
    \includegraphics[trim={8mm 221mm 70mm 10mm}, clip, width=1\columnwidth]{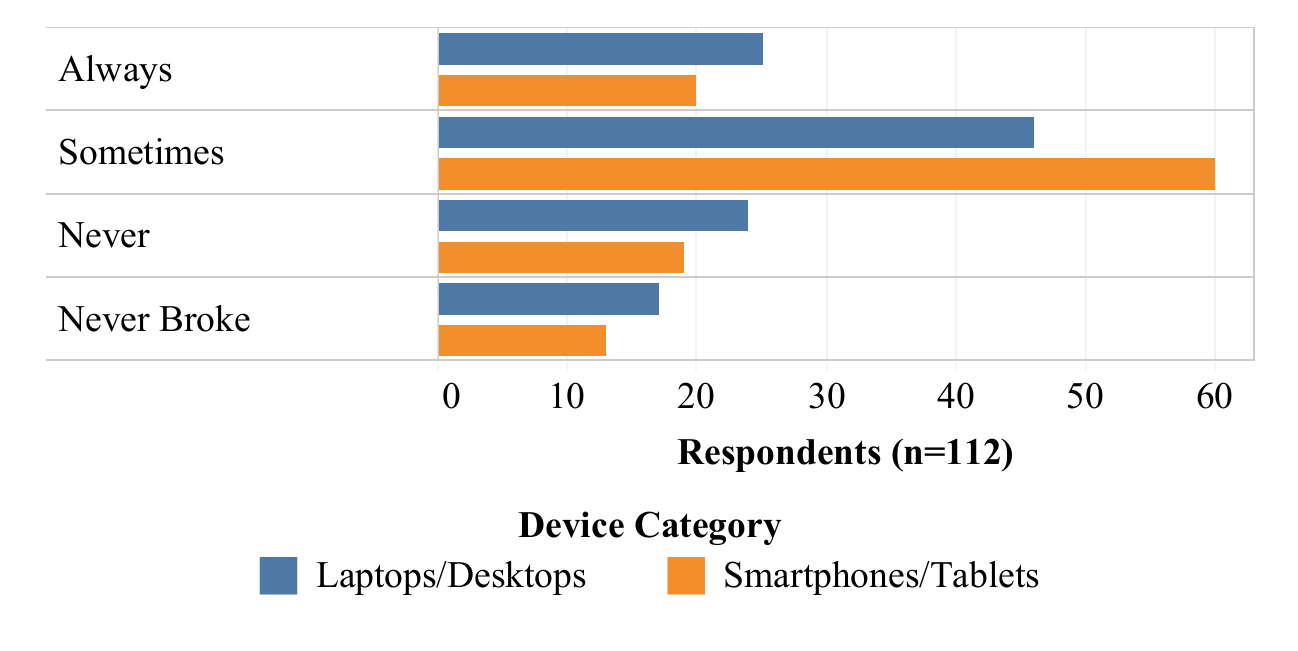}
    \caption{Responses to ``In the last five years, when a personal device broke or required service, how often did you get it repaired or serviced?''}
    \label{fig:surveyRepairLikely}
\end{figure}

First, we established how many devices for each category broke down and if the broken devices were repaired (both commercially or non-commercially) in the past five years (see Figure~\ref{fig:surveyRepairLikely}).
For the smartphones and tablets category, 12\% (13/112) of respondents reported no broken devices,  17\% (19/112) reported never getting repairs with at least one device requiring repair,  54\% (60/112) reported getting devices repaired sometimes, and  18\% (20/112) reported always getting broken devices repaired. 
For the laptops and desktops category, 15\% (17/112) respondents reported no broken devices,  21\% (24/112) reported never getting repairs with at least one device requiring repair,  41\% (46/112) reported getting devices repaired sometimes, and  22\% (25/112) reported always getting broken devices repaired.

For respondents who chose not to get at least one device repaired, we asked all possible factors that influenced their decision not to get the device repaired. 
For  79 cases where broken smartphones/tablets were not repaired, the cost was reported as a factor for 58\% (46/79) cases, the hassle was a factor for 46\% (36/79) cases, the device was not repairable for 33\% (26/79) cases, and privacy was a factor for 33\% (26/79) cases. For one broken device, the owner chose ``other'' as their response and stated that they were getting a new device anyway. 
For 70 cases where broken laptops/desktops were not repaired, the cost was reported as a factor for 57\% (40/70) cases, the hassle was a factor for 34\% (24/70) cases, the device was not repairable for 27\% (19/70) cases, and privacy was a factor for 33\% (23/70) cases. In one case, the device owner chose ``other'' as their response, and their reason was their inability to find a suitable service provider.  In \S~\ref{sec:interview}, we investigate why
privacy was not cited as a more important factor. 

We asked respondents if they got their devices repaired by a commercial service provider. 56\% (63/112) and 46\% (52/112) reported getting a device repaired by a commercial service provider for smartphones/tablets and laptops/desktops categories, respectively. 
A total of 86 respondents reported getting one or more devices repaired at a commercial service provider. 
In terms of where these repairs were performed, 37\% (32/86) reported getting a device repaired at a large electronics retail store, 64\% (55/86) at a kiosk or a small repair store, 14\% (12/86)  at a commercial in-home repair service, 9\% (8/86)  at their organization's IT department, 27\% (23/86) mailed the device to the manufacturer or brought it to a manufacturer store, and 13\% (11/86)  at their cellular service provider. 
In \S~\ref{sec:interview}, we investigate why people chose a particular service provider for repairs.

Finally, we collected information on what type of repair services were performed (multiple repairs were possible for one respondent). For the smartphones/tablets category, the most requested services were screen replacement, battery replacement, and charging system repair, with 62\% (39/63), 56\% (35/63), and 25\% (16/63) of respondents, respectively, reported having requested that service. For the laptops/desktops category, the most requested services were virus removal, battery replacement, and screen repair, with 38\% (20/52), 33\% (17/52), and 29\% (15/52) of respondents, respectively, reported having requested that service. 

\subsubsection{Repair Protocol}
We asked respondents if they were asked to provide or remove their PIN or password for a repair (multiple repairs were possible for one respondent).
Of 63 respondents who got smartphones/tablets repaired, 41\% (26/63) were not asked to provide their PIN or password for any repair. 
14\% (9/63) of respondents did not have authentication enabled on at least one repaired device. 
37\% (23/63) were asked to provide their PIN or password for at least one device, and 17\% (11/63) were asked to remove their PIN or password for at least one device.
Of 52 respondents who got laptops/desktops repaired, 37\% (19/52) were never asked to provide their PIN or password.
8\% (4/52) had a device repaired that was not protected with a PIN or password. 
38\% (20/52) were asked to provide their PIN or password for at least one device, and 27\% (14/52) were asked to remove their PIN or password for at least one device.
\new{These findings do not match those from the first part of the study, where almost all service providers requested credentials. This difference is due to the broader scope of the survey as it included locations where credentials may not be typically requested (14\% of respondents used an in-home repair service, 13\% used their cellular service provider, and 9\% used their organization’s IT department).}

Respondents reported complying with the technician's request.
All respondents provided their PIN or password when requested by the technician. 
We asked respondents if they felt that providing or removing credentials was necessary for the repair. 
20\% (9/46) of these respondents felt that sharing or removing was not required, 24\% (11/46) felt that it was required, while 57\% (26/46) were uncertain whether there was a need to share or remove the PIN or password. 
We also asked  respondents how comfortable they were removing or providing their PIN or password. Their responses were provided on a 5-point Likert scale. Responses were well distributed, with 4\% (2/46) of respondents indicating they were uncomfortable, 39\% (18/46) were somewhat uncomfortable, 7\% (3/46) were neither comfortable nor uncomfortable, 39\% (18/46) were somewhat comfortable, and 11\% (5/46) were comfortable. 
In \S~\ref{sec:interview}, we investigate why a significant number of respondents are surprisingly comfortable sharing or removing their credentials. 


We asked respondents if they were requested by the service provider to remove their personal data before the repair. 69\% (59/86) respondents reported that they were not asked, 26\% (22/86) reported that they were asked to and they did remove data, and 6\% (5/86) reported that they were asked to and they did not remove data. 

\subsubsection{Threats to Personal Data}
We asked respondents how concerned they were regarding the privacy issues surrounding electronics repairs. Their responses were provided on a 5-point Likert scale.
Only 2/86 (2\%) reported being not concerned, 22/86 (26\%) reported being slightly concerned, 22/86 (26\%) reported somewhat concerned, 28/86 (33\%) reported being moderately concerned and 12/86 (14\%) reported being extremely concerned. A Kruskal-Wallis test examined the effect of respondents' technological proficiency on their reported level of concern and found no significant differences (H(2) = 0.39, p = 0.83).



We asked respondents if they felt that their personal data might have been inappropriately accessed during their last repair service. 
Their responses were provided on a 5-point Likert scale.
8/86 (9\%) respondents felt that their data was definitely not accessed, 26/86 (30\%) felt that their data was probably not accessed, 19/86 (22\%) were unsure, 22/86 (26\%) felt that their data was probably accessed, and 11/86 (13\%) felt that their data was definitely accessed.
A Kruskal-Wallis test examined the effect of respondents' technological proficiency on their reported level of likelihood that inappropriate access occurred and found no significant differences (H(2) = 2.57, p = 0.28).



Finally, we asked  respondents if these privacy concerns made them reluctant to get their device repaired. 
50\% (43/86) respondents reported that these concerns did make them reluctant, 34\% (29/86) were not reluctant due to privacy risks, while 16\% (14/86) never considered privacy risks.
A Pearson's Chi-Squared test examined the effect of respondents' technological proficiency on their response and found no significant differences ($\chi^2(4)$ = 4.58, p = 0.33).

\subsubsection{Safeguards for Data Protection}
We asked respondents if they noticed any safeguards implemented by the repair company during the last repair (multiple responses were allowed). 
71\% (61/86) respondents reported not noticing any safeguards.
21\% (18/86) respondents reported that they were informed by an employee, 8\% (7/86) respondents reported that they found this information in the store, and 7\% (6/86) respondents reported that they found this information online. 
If respondents reported noticing any safeguards, we also collected feedback as free-form text on the safeguards that they noticed.
Twenty-two responses were coded, which include removal of hard drive or data (8/22), reassurance by technician (7/22), presentation of terms of service (4/22), and option to get the device repaired in person (3/22). 

Finally, we asked respondents if they took any steps to safeguard their data (multiple responses were allowed). We provided respondents with several options and allowed entry of free-form text.
17/86 (20\%) reported not taking any steps. 
37/86 (43\%) reported backing up and deleting data, 26/86 (30\%) reported deleting some data, 25/86 (29\%) reported changing passwords or logging out of websites, 
9/86 (10\%) ensured that the device or data was encrypted, and 2/86 (5\%) chose ``other'' (they reported that it was impossible for them to do anything).

%% file: 07.Interview.tex
\section{Part-4: Semi-Structured Interview}
\label{sec:interview}

The goal of the semi-structured interview was to get deeper insights into the experiences of participants during the repair process and to get feedback on possible solutions to improve  privacy in the electronics repair industry.
All participants in the interviews had already completed the survey, and we treated the interview as an extension of the survey.\looseness=-1

\subsection{Participants and Procedure}
Respondents who expressed their interest in participating in the follow-up interview and met the inclusion criteria (i.e., had repaired a device across one or more device categories) were invited to participate. 
Eighty-six survey respondents met the inclusion criteria and were contacted over email to participate in the interviews (of which two declined, and 54 did not respond in  time). 

Due to the pandemic, the interviews were conducted online (using Google Hangouts or Microsoft Teams). 
We chose a platform that supported video and screen sharing as it allowed us to share screenshots of the diagnostic feature of the BIOS utility and the logging by Problem Steps Recorder (more details to follow). 
If participants preferred doing the interview over the phone, they were required to have an Internet-connected device to view the screenshots hosted on Qualtrics to provide their feedback.


For participating in the interview, participants were paid \$20. The interview questions were broadly categorized into the following categories and required both categorical and free-form responses (see questions in Appendix~\ref{app:semi_structured_interview}):

\begin{itemize}\itemsep0em
    \item \textbf{Demographic and background:} We collected further demographic information and if they had electronic devices repaired that did not belong to the two categories considered in the survey (e.g., cameras and gaming consoles).
    \item \textbf{Device needing repair:} We asked participants how long they were using the device, if it was shared with someone, if it was encrypted, the personal data that was stored on the device, and the steps the device owner took before dropping off the device.   
    \item \textbf{Choice of service provider:} We asked participants where they got their device repaired, what repair was performed, and why they chose a particular service provider. 
    \item \textbf{Repair experience:} We asked participants about their repair experiences, including the instructions they were provided during the drop-off, any communications regarding their privacy, and any suspicions of privacy violations. 
    \item \textbf{Possible improvements:} We asked participants their thoughts on possible ways the threat to their personal data could be mitigated and collected their feedback on two possible solutions to safeguard the personal data of users during repairs.
\end{itemize}

\subsection{Results}
\label{subsec:interview_results}
For qualitative analysis of participants' responses, two researchers independently performed open coding to identify codes or themes in participant responses to free-response questions.
\new{We calculate and report inter-rater agreement over identified themes using Fleiss' Kappa.}
These researchers then compared and discussed identified codes until a consensus was reached. 
Several other researchers (e.g., Acar et al.~\cite{acar2017internet}) in this domain have used this approach. When reporting results from interviews, we report quotes from participants when they represent a theme. We also report the number of participants who expressed that theme and provide a representative quote. 

The demographic information for participants of the semi-structured interviews is provided in Table~\ref{tab:demographicsInterview}. For the interview participants, annual household income levels and highest education level achieved are reported. The table shows that our participant pool has good diversity for these demographics, which is important for our investigation.

\begin{table}[t]
\caption{Interview Participant Demographics}
\label{tab:demographicsInterview}
\centering
\small\addtolength{\tabcolsep}{-3.75pt} \renewcommand{\arraystretch}{1.05}
\begin{tabular}{cccccccc}
\multicolumn{8}{c}{{\bf n = 30}} \\ \hline \hline
\multicolumn{8}{c}{{\bf Gender} } \\
\multicolumn{4}{c}{\emph{Female}}  & \multicolumn{4}{c}{\emph{Male}}  
 \\
\multicolumn{4}{c}{16}    & \multicolumn{4}{c}{14}     \\ \hline
\multicolumn{8}{c}{{\bf Age (in years)}} \\
\emph{18--25}               & \emph{26--30}               & \emph{31--35}               & \emph{36--40}               & \emph{41--45}               & \emph{46--50}               & \emph{50+}                                \\
11                 & 10                & 5                 & 3                 & 0                 & 0                  & 1                                                   \\ \hline
\multicolumn{8}{c}{{\bf Annual Household Income ($\times$ \$1000)} }  \\
\multicolumn{1}{c}{\emph{\textless{}\$30}}   & \multicolumn{2}{c}{\emph{\$30--74}}             &  \multicolumn{2}{c}{\emph{\$75--99}}              &  \emph{\textgreater{}\$100}  & \emph{Undisclosed} \\
 \multicolumn{1}{c}{2} & \multicolumn{2}{c}{12}  & \multicolumn{2}{c}{6}  & 8 & 2                  \\ \hline
\multicolumn{8}{c}{{\bf Highest Education Level}} \\
\multicolumn{3}{c}{\emph{High School}}                                                       &
\multicolumn{3}{c}{\emph{Undergraduate}}                                  & \multicolumn{2}{c}{\emph{Graduate}}                           \\
\multicolumn{3}{c}{5}                                           & \multicolumn{3}{c}{21}  & \multicolumn{2}{c}{4}                                          \\ \hline
\multicolumn{8}{c}{{\bf Self-Reported Proficiency in Technology}} \\
& \multicolumn{2}{c}{\emph{Basic}}                                         & \multicolumn{2}{c}{\emph{Intermediate}}                                   & \multicolumn{2}{c}{\emph{Advanced}}                            \\
& \multicolumn{2}{c}{2}                                           & \multicolumn{2}{c}{24}                                           & \multicolumn{2}{c}{4}                    &                                   \\  \hline \hline
\end{tabular}
\end{table}

\subsubsection{Broken Devices}
Participants reported having 61 broken smartphones (69\% (42/61) got repaired), 41 broken computers (63\% (26/41) got repaired), and 19 other broken electronics (58\% (11/19) got repaired). Other devices included 14 tablets, two game consoles, two cameras, and one dashcam.
Participants reported using the broken devices on average for 2.6~years (median = 2~years) before those devices broke.
Participants' responses were codified to understand why they would not get some broken devices repaired. \new{The inter-rater agreement between the two researchers was substantial (Fleiss's $\kappa$ = 0.76)} Their responses show that only for 5\% (2/42) instances of broken devices not getting repaired, privacy was the major factor for not getting the device repaired. 
However, comments show that due to the end-of-life of the device (24\% (10/42)) and the cost of the repair (24\% (10/42)), some device repairs were never on the table. Comments were similar to: 
\begin{Q}
\textit{``Phone was old, and I was thinking of getting a new one anyway.''}  (P13)
\end{Q}
\begin{Q}
\textit{``It [smartphone] would have been more expensive to repair than replace.''}  (P9)
\end{Q}


\begin{figure}[t]
    \centering
    \includegraphics[trim={12mm 228mm 80mm 12mm}, clip, width=1\columnwidth]{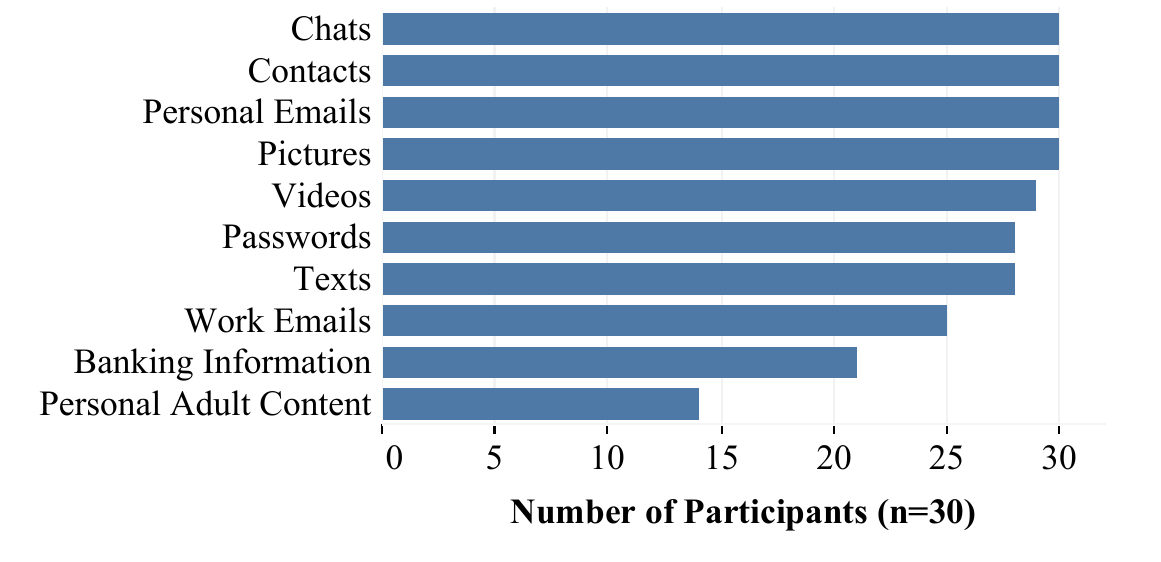}
    \caption{Participants' responses to ``Did the devices you got repaired contain any of the following sensitive data?''}
    \label{fig:surveyDataType}
\end{figure}

The majority of the reported repaired devices (96\% (76/79)) were owned by the participant, while 4\% (3/79) were work devices. Figure~\ref{fig:surveyDataType} shows what type of data was stored on the broken devices. It shows that participants stored all types of sensitive data on their devices. Almost half of the devices also contained personal adult content, which was the target of most privacy violations during the second part of the study.
We asked participants if the device or data was encrypted at the time of the repair. The encryption status was unknown for  80\% (63/79), 1\% (1/79) were encrypted, and 19\% (15/79) were not encrypted.
We also asked participants who they shared each repaired device with. 94\% (74/79) were not shared with anyone, 2\% (2/79) were shared with a spouse, and 4\% (3/79) were shared within the household. 

\subsubsection{Choice of Service Provider}
To build solutions that address privacy concerns, it is important to understand why device owners choose a particular service provider.
For the interviewed participants, 53\% (42/79) devices were brought to a local store, 10\% (8/79) devices were brought to a national store, 33\% (26/79) were brought to the manufacturer, and 4\% (3/79) were brought to a wireless service provider. 
We asked participants the rationale behind why they chose a specific service provider (multiple responses were accepted). Their responses were codified \new{and the inter-rater agreement between the two researchers was almost perfect (Fleiss's $\kappa$ = 0.98). Participants' responses showed} that the reputation of the service provider was one of the main factors in their decision-making process (40\% (12/30)). Their responses indicated that the reputation was perceived due to the brand name of the service provider, the recommendations of others, or Google reviews. We also note that the reputation led to trust:  
 \begin{Q}
\textit{``I trusted them because they were professionals and they had very favourable Google reviews so that builds.''}  (P23)
\end{Q}
\begin{Q}
\textit{``I was ambivalent to get the device repaired, and the big store [a national service provider] seemed like the best bet. Likely because of the reputation and I did not want to get ripped off.''}  (P30)
\end{Q}
Cost and warranty coverage were reported as the other main factors when choosing a service provider (40\% (12/30) and 40\% (12/30), respectively). Their responses were similar to: \textit{``to get the warranty to cover it''} (P10).
Convenience was the third most cited reason (37\% (11/30)). Their responses were similar to: \textit{``It was close to my house and the first place I thought of.''} (P12).
A quick turnaround time was cited as another reason (13\% (4/30)).

 \subsubsection{Risk Consideration}

We asked participants if they considered any risks related to the repair (multiple responses were allowed). 
Their responses were codified \new{and the inter-rater agreement between the two researchers was perfect (Fleiss's $\kappa$ = 1.0). The responses showed} that risks to personal data (30\% (9/30)), work data (7\% (2/30)), and banking information (10\% (3/30)) were most concerning for participants. The responses for personal data were similar to:
\begin{Q}
\textit{``I was worried they would go through my pictures and keep pictures.''} (P7)
\end{Q}

57\% (17/30) participants reported that they did not consider any risks prior to the repair. Their responses show that reputation and trust played a major role in alleviating their concerns. Their responses were similar to: 
\begin{Q}
\textit{``I just didn't think of it, I thought [national service provider redacted] was a big place and safe.''} (P3)
\end{Q}
\begin{Q}
\textit{``It was for my phone but I trusted [manufacturer redacted] enough''} (P17)
\end{Q}

We asked participants what data categories they were most concerned about (multiple responses were allowed). 
Most participants were concerned about their banking information (87\% (26/30)), 57\% (17/30) were concerned about their passwords, 40\% (12/30) were concerned about their emails, and 40\% (12/30) were concerned about personal adult content. 
When asked what the technician would do with their data, 53\% (16/30) reported use for financial gains, 9\% (5/30) reported identity theft, and 13\% (4/30) reported online account takeover. In the second part of the study, we found prevalent casual snooping. During the interview, only 9\% (5/30) of participants reported that technicians would casually snoop on their data and possibly share. Their responses were similar to:
\begin{Q}
\textit{``They wouldn't do anything but just snoop.''} (P16)
\end{Q}
\begin{Q}
\textit{``They could share pictures among themselves.''} (P29)
\end{Q}

\subsubsection{Repair Experience} 
We present participants' repair experiences separately before, during, and after the repair.
For 79 reported repair cases, the most requested repairs were screen repairs (23\% (18/79)) and general hardware repairs (23\% (18/79)), followed by battery replacement (18\% (14/79)) and operating system issues (16\% (13/79)).

\subhead{Preparing for repairs} We asked participants how they prepared their devices before dropping them off for repair on their own (multiple responses were possible). 
50\% (15/30) of participants reported doing nothing on their own. 20\% (3/15) of these participants did not do anything because they did not share the credentials to the locked device. 17\% (5/30) of participants did not do anything because they could not access the device's OS.  13\% (4/30) of participants deleted their data, and 13\% (4/30) backed up their device prior to the repair. 10\% (3/30) reported removing their credentials, so they would not have to provide them or remove them at the repair location. 
We noted that four participants who reported taking some actions to protect their data may not have done it properly. Their responses were similar to:

\begin{Q}
\textit{``I moved pictures into a file folder and relabelled [renamed] it.''} (P6)
\end{Q}

\begin{Q}
\textit{``I uploaded all my photo gallery to Google Photos and then signed out on the device. I was still so scared because all my banking still auto-completes, and I felt sketchy giving it to them.''} (P29)
\end{Q}

For 87\% (69/79) of repairs, participants reported that the service providers did not provide them with any instructions on what to do with the device before handing it over to them. For the remaining repairs, participants were instructed to backup the device (6\% (5/79)), disable ``Find My iPhone'' (2\% (2/79)), remove the SIM card (1\% (1/79)), charge the battery (1\% (1/79)), or factory reset the device (1\% (1/79)). 

\subhead{Interactions with service providers}
During the interview, participants reported not seeing any privacy notices or communications from service providers on controls to safeguard their data.
Half of the participants did recall signing terms of service that include statements about privacy. Their responses were similar to: 
\begin{Q}
\textit{``I signed something that they will withhold confidentiality of the passwords you give them.''} (P23)
\end{Q}
Three participants reported asking about safeguards to protect their data. One participant was provided no answer (\textit{``I did ask, but he [technician] had no answer.'')} (P1). One participant was offered that they could watch the repair in the backroom. One participant reported:
\begin{Q}
\textit{``They [service provider] mentioned they were trained, and they had liability insurance.''} (P5)
\end{Q}
17\% (5/30) of participants reported that they were not asked to provide their credential or remove it for the repair. Only 25\% (4/16) of participants who were required to provide their credentials thought that the technician had a plausible reason to request the credentials. While only one participant reported being comfortable sharing their credentials, all participants reported sharing their credentials. Their reason for sharing was the lack of options. The responses were similar to:
\begin{Q}
\textit{``I did it because I didn't have money for a new phone and there was no other option.''} (P7)
\end{Q}
Participants who were uncomfortable sharing or removing their credentials did not voice their concerns to avoid the technicians scrutinizing their device:
\begin{Q}
\textit{``I was not comfortable removing it, but I knew they would ask me to give the PIN so I pretended that there was nothing on the device.''} (P30)
\end{Q}


\subhead{Post repair}
We asked participants if they noticed anything that indicated that their privacy has been compromised.
17\% (5/30) participants reported that they did not know how to find such evidence and 83\% (25/30) reported not finding any evidence. 
Note that our findings from the second part show that service providers often clear their snooping activity tracks.
 7\% (2/30) participants reported that they changed their passwords after a device repair. 

\subsubsection{Possible Improvements}
\begin{figure}[t]
    \centering
    \includegraphics[trim={14mm 252mm 80mm 12mm}, clip, width=1\columnwidth]{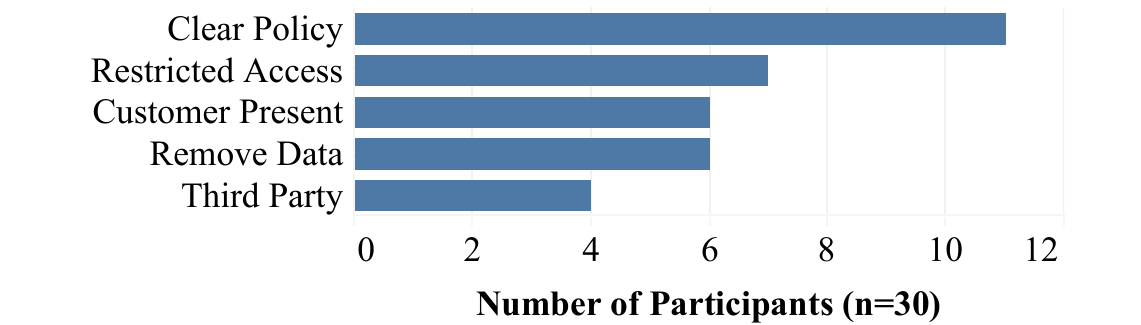}
    \caption{Participants' codified responses to ``Could you suggest a change to the repair process that would make you more comfortable with getting your electronic devices repaired?''}
    \label{fig:interviewImprovements}
\end{figure}
We first asked participants what change could be introduced to better protect their data during repairs (multiple responses were accepted). 
The codified responses (see Figure~\ref{fig:interviewImprovements}) show that 37\% (11/30) of participants desired clear and consistent policies on access to customers' data. \new{The inter-rater agreement between the two researchers was perfect (Fleiss's $\kappa$ = 1.0).} Their responses were similar to:
\begin{Q}
\textit{``Guarantee or policy that is public that they don't go through your device... or audit page after the repair showing what they accessed and why''} (P5)
\end{Q}
\begin{Q}
\textit{``If they don't need my password, don't ask for it. [For the same issue] [Service provider redacted] asked for my password and  [Service provider redacted] didn't, so clearly you can do it without my password. He [technician] shouldn't have asked for it.''} (P8)
\end{Q}
23\% (7/30) reported a method of restricting access to personal data not required for the repair (\textit{``If you could give them a temporary password that they could use for repair''} (P21)). 20\% (6/30) proposed that repairs should be done with customers present, and 20\% (6/30) suggested customers should remove their own personal data before a repair.
13\% (4/30) reported the need for a third party. Their responses were similar to: 
\begin{Q}
\textit{``Some sort of third party that gave you assurances. Sort of like PayPal with repairs.''} (P1)
\end{Q}
\begin{Q}
\textit{``A governing body to oversee the repair industry. Especially with kiosks and small stores.''} (P19)
\end{Q}

 
We also sought feedback on two possible solutions that work for limited scenarios (see \S~\ref{sec:discussion} for limitations of possible solutions). 
First, some frequent repairs could be diagnosed and verified using the diagnostic utility provided by the device manufacturer and without accessing the device OS (e.g., screen, battery, system board). 
Second, we informed participants about a logging utility that would record all the steps taken by the repair technician on their device in the same format as the Problem Steps Recorder. We informed participants that while technicians could tamper with the log, the owner would know if the logs were tampered with, which may indicate a privacy violation. 
For both solutions, we asked participants if these solutions would be effective in protecting their personal data during (applicable) repairs and whether participants would use these solutions for device repair services. For the diagnostic utility, we also asked participants if they were aware of this utility.  
 
73\% (22/30) of participants were not aware of the diagnostic utility. 
28/30 (93\%) and 28/30 (93\%) participants reported that the solution was effective or very effective at preventing a potential privacy breach for the diagnostic and logging utility, respectively.
1/30 (3\%) and 1/30 (3\%) 
participants reported that the solution was somewhat or not effective for the diagnostic and logging utility, respectively. We observed the same pattern for possible adoption---all but one participant wanted to use (or try) these solutions. The one participant's concern with the logging utility was:
\begin{Q}
\textit{``It's all after the fact [violation] so it doesn't have an effect. It would be better if I could see it in real-time than after the fact.''} (P16)
\end{Q}



%% file: 08.Discussion.tex
\section{Discussion}
\label{sec:discussion}

\subsection{Revisiting Research Questions}
\label{sec:revisiting_RQ}
\subhead{RQ1: Existence and communication of privacy policies} We observed that while some service providers shared a privacy policy, it was a generic policy on data collection from customers during retail transactions. These policies did not address key questions for the device repair use case, such as how long user data (such as backups and credentials) are stored, who has access to it, and what controls are in place to protect customers' privacy. Furthermore, despite the media highlighting several privacy violations in this space, no satisfactory response was provided when the investigator asked about controls in place to protect customers' data. We note the possibility that some service providers may have implemented some controls or measures, but the customers were oblivious to them. 

\subhead{RQ2: Level of access requested} Our investigation shows that the service providers generally demand more access than required for the requested repair. 
This demand for greater access is not necessarily driven by their desire to snoop but for efficiency and productivity reasons, as evident by the documentation from the manufacturers (see \S~\ref{subsec:part1_results}). 
However, even when the repair was possible without providing the credentials, most service providers would not make an exception.
\new{Service providers want to repair more devices in the same amount of time using a consistent and manageable process. The difficulty in predicting which component is the source of the problem makes it difficult to determine the type of access required.
Consequently, they prefer complete access to the device for a speedy diagnosis, repair, and verification without having to contact the user again~\cite{limoncelli2016practice}. In \S~\ref{subsec:mitigations}, we discuss the need for a balance between efficiency and privacy.}
The technicians who received the device at the national service providers stated that they were unable to take the device without the credentials. 
If the technicians were truthful about this process, it indicates a policy failure; otherwise, it is a process failure.
 
\subhead{RQ3: Occurrence and types of privacy violations} The second part of our study showed that while data theft was uncommon, casual snooping of customers' data was a regular occurrence. Accordingly, viewing of (revealing) pictures or casual folder snooping was noted as the most common violation. We note that while our logs did not provide any evidence of the theft of financial data, the technicians may have copied it using other means (i.e., copied to a paper). Our investigation also showed that the technicians made an effort not to leave a trace of their snooping. Evidence of snooping on confidential information has been previously reported in other domains, including the workplace~\cite{prestwood2016snooping} and by medical office workers~\cite{boxwala2011suspicious}. Research into why workers snoop is limited but boredom, curiosity, entitlement, and being treated unfairly have been suggested~\cite{prestwood2016snooping}. The reasons for snooping in the repair industry should be established for the design of effective policies and controls to prevent snooping.\looseness=-1 

\subhead{RQ4: Risk perceptions and strategies of customers} The survey respondents chose not to get 33\% of their broken devices repaired because of data privacy concerns. 
For participants who got the device repaired, privacy was a concern for the majority. However, due to the broken device's essential nature or the fear of losing their data, participants proceeded with the repairs.
When taking their device in for repair, most participants were concerned about threats to their financial data or identity.
In terms of safeguarding their data, participants were often unable to adequately protect their data because they could not access the device's OS.
Technologically proficient participants adopted various successful strategies, 
but some participants adopted less effective measures due to their lack of knowledge of the controls available to them, including hiding files or pretending that they did not have anything to hide.

\subhead{RQ5: Potential solutions} 
The interview provided us with insights on various fronts, which are discussed in \S~\ref{subsec:mitigations}. 

\subsection{Reputation, Trust, and Privacy}
\label{subsec:reputation}
Our interview shows that participants' risk preferences are influenced by the reputation of the service provider.
Participants placed trust in service providers that had a better reputation, which lowered participants' perceived risks. 
The reputation of the service provider was often cited as a deciding factor when choosing a service provider. 

The reputation of an entity has been attributed to specific attitudes and capacities~\cite{golberg2001shostack}.
In the context of electronics repair, it is the service providers' capacity to adequately perform the repair  (i.e., their technical competence). 
However, reputation leads to a cognitive bias, where the customer trusts the service provider with their data based on the technical competence of the service provider. 
The implicit assumption is that the repair will be performed in an appropriate way (i.e., by respecting existing conventions and regarding the customer's interests and privacy).
The documented cases of privacy breaches by all types of service providers emphasize the need to distinguish between well-reputed and trustworthy.

Service providers' trustworthiness with customers' data can be established through different sources, including legal rules, security controls, open policies and processes, and professional ethics~\cite{sartor2006privacy}. 
Not all measures are foolproof---legal proceedings require evidence and are uncertain and costly; security controls can be bypassed, and professional ethics give misleading indications. 
\new{Furthermore, 
instances of abuse of customers’ data have been reported for the employees of other types of service providers, including Facebook~\cite{facebook_breach} and Google~\cite{google_breach}. While useful, the existing code of ethics for individuals who handle such data (e.g., the USENIX System Administrators' Code of Ethics~\cite{lisa_ethics}) needs to be complemented. More research is needed for a framework that relies less on trust and more on procedures and controls that prevent privacy breaches when employees handle customers’ data.
}



\subsection{Data Protection in the Repairs Industry}
\label{subsec:mitigations}
The repair service industry saves consumers' the cost of buying a new device and reduces electronic waste.
However, protecting customers' data in the repair industry is a challenging problem.
Customers often have only basic familiarity with technology and are unaware of the controls that are available to them to protect their data.
Different devices may have different controls available to protect data (e.g., folder encryption on computers). 
Different devices or manufacturers may provide diagnostic coverage across different components.
Furthermore, on a broken device, security controls (e.g., data encryption) may not be possible.
\new{On the other hand, service providers want an efficient and consistent process. 
However, efficiency and consistency must be balanced with the potential for privacy violations as} technicians are human and subject to curiosity, which leads to snooping.
We believe that one control or one stakeholder may not be able to solve this problem in a satisfactory way. 
A reliable solution requires actions from three stakeholders---device manufacturers and OS developers, service providers, and regulatory agencies.

Device manufacturers \new{can take a more proactive approach by} standardizing the diagnostic interface to minimize manufacturer-specific differences. Some devices come with the ability to create guest accounts, which provide limited access to the device for some repairs. \new{Samsung has recently introduced a ``repair mode'' to their smartphones~\cite{amadeo2022repair}. Technical details are limited but once this feature is enabled, only default apps can be accessed and access to user data is blocked until the credentials are provided.}  
OS developers \new{can provide some reactive controls such as} a tamper-resistant logging utility. 
Schneier and Kelsey proposed a cryptographic mechanism to secure logs on an untrusted machine~\cite{schneier1998cryptographic}.
Similarly, researchers have leveraged modern technologies, such as Intel SGX and trusted execution, for tamper-resistant logging~\cite{paccagnella2020custos, karande2017sgx}.
The Problem Steps Recorder logging format was found to be ``easy to follow'' by the participants as it ``showed'' them everything that was accessed. 
While such a logging utility will catch or signal possible privacy violations, service providers may not be willing to disclose their repair techniques. 
These solutions are not perfect---guest accounts or the diagnostic utility are not applicable for repairs like virus removal, and enabling tamper-resistant logging or ``repair mode'' requires a functioning device. 
\new{Furthermore, these controls would require users to enable them, and users may not be aware of these features or forget to enable them.}
However, through open policies, service providers can guide customers on the best control for the device condition and requested repair. 

Service providers need to define a policy and adopt controls that could be used to protect  customers' data from malicious technicians.
In terms of policy, while less efficient, service providers must move beyond the ``all-access'' model. Several common repairs can be completed using the device manufacturers' diagnostic utility and others through a guest account on the device (e.g., screen repair or device battery). 
Service providers should communicate why a specific type of access is requested and the technical controls that customers can use to protect different types of data.
Ideally, such information should be made available on the website by service providers or communicated by device manufacturers in their mail-in instructions so that customers can complete those tasks before handing over their device.
Finally, service assessment has been a part of other service industries, which is achieved through non-technical means such as video recording with random audits or mystery shoppers. 
Such service assessment should be a part of the repair service industry.
It is critical that such policies should be consistent across different types of service providers, which can be ensured through a regulatory agency.

Regulatory bodies need to play a strong role in safeguarding the privacy of consumers in the repair industry.
The mandate of the Bureau of Consumer Protection of the US Federal Trade Commission includes the protection of consumers against unfair, deceptive, or fraudulent practices and the education of consumers~\cite{ftc}. 
Similarly, Section 18 of the Personal Information Protection and Electronic Documents Act empowers the Canadian Office of the Privacy Commissioner to undertake an audit of the personal information management practices of an organization~\cite{opc_report}. 
Other agencies have set precedents in their respective areas. 
For instance, in the US, under the Federal Food, Drug, and Cosmetic Act, the Federal Food and Drug Administration (FDA) has provided guidelines for Public Health to routinely inspect food service establishments to ensure proper processes are being followed~\cite{fda}.
A certificate at each establishment communicates the results to the customers to make informed decisions.
A similar strategy needs to be adopted for the electronics repair industry to ensure that privacy breaches are no longer common.

%

%% file: 09.Limitation.tex
\section{Limitation}
\label{sec:limitation}

Similar to other studies involving human participants, our study has several acceptable limitations. 
Our online survey and semi-structured interview contain self-reported data, which may be influenced by several factors, including the participants' memory, understanding, or subjective views. Participants may have misreported aspects in an effort to avoid embarrassment or to provide favourable responses to the researchers. 
Since these limitations are not avoidable, we focus on the aspects that were specific to our study.

During the first part, we assumed that the battery replacement would not require access to the device OS. While one service provider agreed to do it without the credentials, technicians at other service providers may have erroneously believed that they needed credentials for diagnosis or repair. 
It is possible that some service providers may have implemented controls to protect customers' data but the technicians did not inform us about those controls.
For the second part, our sample size is small. Furthermore, it is subject to several other limiting factors that are not in our control, including a curious technician not finding the opportunity to snoop due to some reason or the presence (or absence) of privacy violations indicating that a service provider regularly commits (or does not commit) such violations.  

There was a six-week gap between the participants' online survey responses and their semi-structured interviews. This gap and the questions posed in the survey may have influenced participants' responses during the interview.
During the online survey and the interview, participants reported their experiences that may have occurred in the past five years.  As such, their recall of the details  may not have been accurate. However, the experiences reported by the participants closely resemble our recent experiences, which show an absence of policies and protocols to prevent access to customers' data.

%% file: 10.Conclusion.tex
\section{Conclusion}
\label{sec:conclusion}

We conduct the first-ever comprehensive study to understand the state of privacy in the electronic repair services industry. 
Our investigation shows an absence of policies and controls to safeguard customers' data across all types of repair service providers. 
We show that despite their concerns, customers get their devices repaired to save cost or data.
We provide suggestions on how the state of privacy in the repair industry can be improved. 
Our work calls to action device manufacturers, OS developers, repair service providers, and regulatory bodies to take appropriate measures to safeguard customers' privacy in the repair industry.

%% file: 11.Apendix.tex
\appendix

\section{Appendices}
\subsection{Phase-1 Survey for Researchers}
\label{app:reconnaissance_survey}
The following questions were completed by the researchers within five minutes of leaving the repair store. \\
\textbf{Demographics and Background.}
\begin{enumerate}
\item[1] Name and Location of Repair Center: 
\item[2] Date and Time of Visit: 
\item[3] Researcher Name: 
\item[4] Researcher Gender: 
\item[5]Device Type:  
\item[6] Did you see any notices/policies on repair privacy? 
\item[7]What repair was the device brought in for: 
\item[8] Did the technician request any of the following: \\
(a) Provide PIN or other authentication credentials; (b) Remove PIN or other authentication; (c) Remove data, unlink/remove accounts or factory reset
\end{enumerate}
\textbf{Provide PIN or other Authentication Credentials, Remove PIN or other Authentication, Remove data, unlink/remove accounts or factory reset}\\
Answer the following questions for each of the above you were asked to perform.
\begin{enumerate}
\item[9] What were you asked to do?
\item[10] Response to ``Why is this required?'' 
\item[11] Response to ``Can you do it without doing the action?'' (if not answered by previous response) 
\end{enumerate}
\textbf{Additional Questions.}
\begin{enumerate}
\item[12] Response to  ``Will anyone access my data during the repair?'' 
\item[13] Response to ``How do you ensure no one accesses my data during the repair?'' 
\end{enumerate}



\subsection{Online Survey}
\label{app:online_survey}
\textbf{Demographic Information.}
\begin{enumerate}
\item[1] How old are you? \\
(a) 18-25 years old; (b) 26-30 years old; (c) 31-35 years old; (d) 36-40 years old; (e) 41-45 years old; (f) 46-50 years old; (g) over 50 years old; (h) choose not to respond
\item[2] How do you identify?\\
(a) Female; (b) Male; (c) My gender identity is not listed above; (d) Choose not to respond
\item[3] Which of the following best describes your level of proficiency with technology like smartphones or laptops?\\
(a) Basic (I can perform basic tasks such as sending emails or browsing the internet); (b) Intermediate (I can perform intermediate tasks such as changing the settings or installing new applications); (c) Advanced (I am capable of writing source code)
\end{enumerate}
\textbf{Personal Device Repair.}\\
For the purposes of this survey, personal devices include: Laptop or Desktop Computers, Smartphones (iPhones, Android, etc.), and Tablets or tablet computers. \textbf{Repairs} on personal devices refer to any of the following services performed by a business such as a retailer, manufacturer, store, service centre or kiosk \textbf{(Repair Business)}: \textit{(list removed due to space constraints)}


\begin{enumerate}
    \item[5]In the last 5 years, when a personal device of each type broke (experienced damage, a fault or malfunction) in any way, how often did you attempt to get it repaired? [Asked for each of the following categories:  Smartphone(s) or Tablet(s) and Laptop or Desktop Computer(s)]\\
    (a)I have not had any devices of this type that broke in the last 5 years.; (b) Never; (c) Sometimes; (d) Always
\end{enumerate}
\begin{enumerate}
    \item[6]\textbf{[IF Always or Sometimes for Laptop or Desktop Computer(s)]} In the last 5 years, for each of these device types, which of the following factors influenced your decision not to get a broken device repaired? (check all that apply.)[Asked for each of the following categories:  Smartphone(s) or Tablet(s) and Laptop or Desktop Computer(s)]\\
    (a) I repaired all devices of this type when they were broken.; (b) Not worth the cost; (c) Privacy concerns; (d) Not worth the hassle; (e) Device was not repairable; (f) Other (\textit{A box for free form text input is provided})
    \item[7] In the last 5 years, have you had any of the following electronic devices repaired by a store, repair centre, kiosk or other commercial device repair service? (Check all that apply)\\
    (a) Smartphone(s) or Tablet(s); (b) Laptop or Desktop Computer(s)
    \item[8] Where were the device(s) selected above brought for repair? (Check all that apply, list any other locations)\\
    (a) Large electronics retailer/store; (b) Repair kiosk / small repair store; (c) Manufacturer (in-store or mailed in); (d) IT department of my workplace, organization or school; (e) Cellular service provider; (f) In-home repair; (g) Other (\textit{A box for free form text input is provided})
    \item[9] What type(s) of repairs were performed on your device(s) selected above? (Check all that apply. Choose N/A if you didn't select a device of that type above.)[Asked for each of the following categories:  Smartphone(s) or Tablet(s) and Laptop or Desktop Computer(s)]\\
    (a) N/A; (b) Screen replacement; (c) Camera repair or replacement; (d) Battery replacement; (e) Charging system repair; (f) Motherboard/logic board repair/replacement; (g) Cellular, Wi-Fi, or Bluetooth repair; (h) Audio/microphone repair; (i) Forgotten passcode/PIN; (j) Water damage repair; (k) Virus/Malware removal; (l) Software installation; (m) Other hardware repair; (n) Other software repair; (o) Unknown repair (\textit{A box for free form text input is provided})
\end{enumerate}
\begin{enumerate}
    \item[10] Were you \textbf{ASKED} to provide or remove your password/login/PIN for any repair for the following device(s) selected above? (Check all the apply)[Asked for each of the following categories:  Smartphone(s) or Tablet(s) and Laptop or Desktop Computer(s)]\\
    (a) N/A; (b) No, I was not asked for my password/login/PIN for any of the repairs; (c) One or more devices were not protected by a password/login/PIN; (d) Yes, I was asked for my password, login or PIN for at least one repair; (e) Yes, I was asked to remove my password, login or PIN prior to at least one repair
\end{enumerate}
\begin{enumerate}
    \item[11]\textbf{[IF asked for PIN/password during a repair]} You answered ``Yes'' for at least one device in the previous question. Did you EVER \textbf{PROVIDE} your password/login/PIN when asked?[Asked for each of the following categories:  Smartphone(s) or Tablet(s) and Laptop or Desktop Computer(s)]\\
    (a) N/A; (b) yes; (c) no
\end{enumerate}
\begin{enumerate}
    \item[12]\textbf{[IF asked for or to remove PIN/password during a repair]} You previously answered that you were asked to provide or remove your password/login/PIN for a repair. For any of these repairs, do you feel that providing or removing your password/login/PIN was not required for the repair?\\
    (a) Yes; (b) Maybe; (c) No
    \item[13]\textbf{[IF asked for or to remove PIN/password during a repair]} You previously answered that you were asked to provide or remove your password/login/PIN for a repair. How comfortable are you doing this for a repair?\\
    (a) Comfortable; (b) Somewhat comfortable; (c) Neither comfortable nor uncomfortable; (d) Somewhat uncomfortable; (e) Uncomfortable
    \item[14] Have you \textbf{EVER} been asked by a repair company to remove your personal data from a device before bringing it in for repair (including factory reset, erase all content, etc.)? (If you have been asked for multiple repairs, answer for the most recent)\\
    (a) No, I was never asked to remove my personal data; (b) Yes, and I removed my data prior to the repair; (c) Yes, and I did not remove my personal data prior to the repair (if you unlinked an account from your device but did not remove your personal data select this option)
    \item[15] Prior to getting a device repaired, have you ever taken any of the following steps to safeguard your personal data? (Check all that apply. Do not select actions you were instructed to do by your repair company.)\\
    (a) I made a copy on a backup device and deleted all the data off the device I was getting repaired.; (b) I deleted some of my data.; (c) I set a password/PIN or set a new password/PIN.; (d) I changed passwords or logged out of websites or apps.; (e) I encrypted the device or ensured it was already encrypted.; (f) Other \textit{(A box for free form text input is provided)}
    \item[16] During the last 5 years, were you ever concerned that your personal information could have been compromised during a repair?\\
    (a) Yes; (b) No; (c) I don't know
    \item[17] How concerned are you with the privacy issues surrounding electronics repair?\\
    (a) Extremely concerned; (b) Moderately concerned; (C) Somewhat concerned; (d) Slightly concerned; (e) Not at all concerned
    \item[18] Has the risk of your personal data being accessed by an unauthorized party ever made you reluctant to get a device repaired?\\
    (a) Yes; (b) No; (c) I never considered this risk
\end{enumerate}
\textbf{Protecting Private Data.}
\begin{enumerate}
    \item[19] At any time during your LAST repair transaction, were you aware of any safeguards implemented by the repair company to protect your personal data? (check all that apply)\\
    (a) Yes, I was informed about one or more measures the repair company was taking to protect my privacy by an employee; (b) Yes, I saw information about one or more measures the repair company was taking to protect my privacy displayed in the store; (c) Yes, I saw information about one or more measures the repair company was taking to protect my privacy online; (d) No; (e) Other (\textit{A box for free form text input is provided}); (f) Not applicable
    \item[20]\textbf{[IF saw or was informed about measures to protect privacy]}You previously answered that you were aware of steps taken by the repair company to protect your privacy. What steps did the repair company take to protect your personal data? (\textit{A box for free form text input is provided})
    \item[21] During your LAST repair, do you feel your personal data could have been inappropriately accessed?\\
    (a) Definitely yes; (b) Probably yes; (c) Unsure; (d) Probably not; (e) Definitely not
\end{enumerate}
\subsection{Semi-Structured Interview}
\label{app:semi_structured_interview}
\textbf{Demographics Information.}
\begin{enumerate}
    \item[1] How old are you? (same options as survey Q1)
    \item[2] How do you identify? (same options as survey Q2)
    \item[3] What is the highest level of education you have achieved?\\
    (a) Some high school; (b) High school; (c) Some college/university; (d) Trade/technical/vocational training; (e) Associate's degree; (f) Bachelor's degree; (g) Master's degree; (h) Professional degree; (i) Doctorate; (j) Prefer not to say
    \item[4] What’s your annual household income?\\
    (a) Under \$15,000; (b) \$15,000-\$29,999; (c) \$30,000-\$49,999; (d) \$50,000-\$74,999; (e) \$75,000-\$99,999; (f) \$100,000-\$150,000; (g) Over \$150,000; (h) Prefer not to answer
    \item[5] Which of the following best describes your educational background or job field?\\
    (a) I have an education in, or work in, the field of computer science, computer engineering or IT; (b) I do not have an education in, nor do I work in, the field of computer science, computer engineering or IT; (c) Prefer not to say
    \item[6] Which of the following best describes your level of proficiency with technology like smartphones or laptops? (same options as survey Q3)
\end{enumerate}
\textbf{Experience Getting Devices Repaired.}
\begin{enumerate}
    \item[7] Please count the number of the following types of devices that required repair/service, and how many you got repaired in each category:[Asked for the following categories: Smartphones, Laptop or Desktop Computers, Tablets, Cameras, Game Consoles]
    \item[8] \textbf{IF device wasn't repaired} For each device that had broken but was not repaired, why was it not repaired?
\end{enumerate}
\textbf{Repaired Device.}
These questions were asked for each device that was indicated as having been repaired.
\begin{enumerate}
    \item[9] What device were you getting repaired? [Researchers noted down the make/manufacture and model. Additionally, Researchers noted down how long its been since the device was repaired.]
    \item[10] Approximately how long have you used the device (or how long have you owned the device)?
    \item[11] Was the device owned by you, Owned by your company, workplace, institution, or school – Only or primarily for work use, or Owned by your company, workplace, institution, or school – Used for both personal and work use?
    \item[12] Did this device contain any of the following sensitive data?[Asked for the following categories: Contacts, Personal Emails, Passwords/Logins Including Apps, Work Emails, Texts, Chats, Pictures, Videos, Personal Adult Content, Commercial Adult Content, Banking Information, Digital Wallet]
    \item[13] Was this your primary device, and was it shared with someone else (spouse, children, etc.)?
    \item[14] Is the device encrypted, or has it ever been encrypted? [Researchers noted responses that had to do with the repair in question 18.]
    \item[15] Where was the device brought for repair (Store, Kiosk, mailed in, manufacturer, IT department, etc.)? Or was it an in-home repair (record if in-home repair service for manufacturer, store, etc.)? What was the rationale behind using this repair vendor? 
    \item[16] What repair or service was performed on the device? Were you aware of the cause of the problem with the device prior to the repair?
    \item[17] Prior to the repair, were you asked to take any steps? If yes, do you know what would happen if the device was too damaged to comply with these instructions?
    \item[18] What steps did you take on your own to prepare the device for the repair (not because you were instructed to by the repair company)? 
    \item[19] Did you consider risks to your personal data when getting the device repaired? If so, what were those risks? Did you have any concerns about your data (name types)? [If no risks were reported, the researcher tried to understand why it wasn't considered.]
    \item[20] At any time during the repair transaction, were you made aware of any safeguards implemented by the repair company to protect your data?
    \item[21] Were you required to provide the repair technician with your PIN, password, or other authentication information? if yes:
    \begin{enumerate}
        \item[21.1] Was a reason given for why it was required?
        \item[21.2] Did the reason seem plausible to you?
        \item[21.3] Were you comfortable giving your PIN/password?
    \end{enumerate}
    \item[22] Did you feel your personal data was safe when getting your device repaired? (\textit{5-point Likert scale ``1'' - ``5, 5 being very safe''})
    \item[23] Do you recall an incident where your personal data may have been compromised during a repair? 
    If yes:
    \begin{enumerate}
        \item[23.1] When did this occur?
        \item[23.2] What makes you believe your data may have been compromised?
        \item[23.3] Did you take any recourse?
    \end{enumerate}
\end{enumerate}
\textbf{General Questions.}
\begin{enumerate}
    \item[24]Which of the following types of personal data are you most concerned about a repair technician accessing: [Researchers recorded free response answers for all categories in Q12.]
    \item[25] If your personal data was compromised during an electronic device repair, what do you believe would be done with the data? 
    \item[26] Could you suggest a change to the repair process that would make you more comfortable with getting your electronic devices repaired?
    \item[27]If there was a way to better protect your personal data during the repair of electronic devices, do you: 
    \begin{enumerate}
        \item[27.1] Feel you would be more likely to use electronics repair services?
        \item[27.2] Be more comfortable during a repair?
    \end{enumerate}
    \item[28] Researchers are proposing a logging program that cannot be disabled by the repair technician. It would be clear if the logs were deleted by the technicians. The following are pictures are examples of the screenshots that would be taken by the program during the repair so you are able to verify your privacy was maintained during the repair. [Researcher showed example output of the Windows Problem Step Recorder.]
    \begin{enumerate}
        \item[28.1] How effective do you feel this proposed logging program would be at preventing privacy breaches?(\textit{5-point Likert scale ``1'' - ``5, 5 being very safe''}) [Researchers also recorded free text response]
        \item[28.2] Would you feel your personal data is safe when getting your device repaired if the logging program was used? (\textit{5-point Likert scale ``1'' - ``5, 5 being very safe''}) [Researchers also recorded free text response]
        \item[28.3] Would you use this solution for repair services if it was available? [Researchers asked why for no and maybe responses] \\
        (a) yes; (b) no; (c) maybe
    \end{enumerate}
    \item[29] [Researchers showed an example of UEFI diagnostics from a HP Laptop (see Figure \ref{fig:hp_bios}).] In this mode, most hardware on the device can be tested but no personal data can be seen or retrieved on the device. The “Diagnostic mode” can be accessed without a PIN/password. This mode exists for most laptops but most tablets and smartphones do not have this feature.
    \begin{figure}[ht]
    \centering
    \includegraphics[ trim={0mm 0mm 0mm 0mm}, clip,width=0.8\columnwidth]{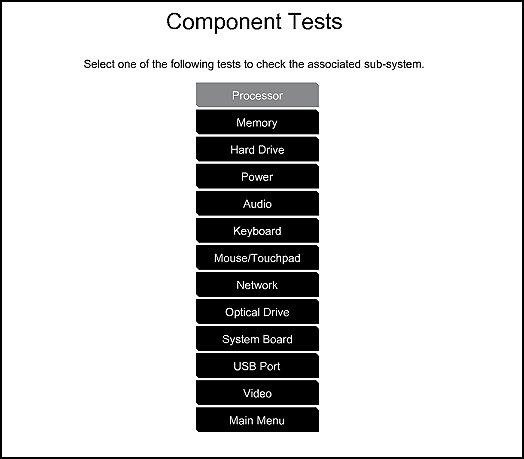}
    \caption{Tests available through HP BIOS Diagnostic Utility}
    \label{fig:hp_bios}
    \end{figure}
    \begin{enumerate}
        \item[29.1] Did you know such a mode existed on laptops for diagnostics? Were you ever made aware of this by a repair facility?
        \item[29.2] Would you be less concerned about your data privacy if a ``Diagnostic Mode”  option existed for the service you are requesting? 
        \item[29.3] Would you be more likely to get your electronic devices repaired if it had this feature?
    \end{enumerate}
\end{enumerate}